\documentclass[aps,prl,twocolumn,groupedaddress,showpacs]{revtex4}
\usepackage{amsfonts}
\usepackage{mathrsfs}
\usepackage{graphicx}
\usepackage{amsmath}
\usepackage{amssymb}
\usepackage{amsbsy}
\usepackage{bm}

\def\bs{\bm{\sigma}}
\def\bx{\bm{\xi}}

\def\hb{\tilde{\beta}}
\def\hG{\tilde{G}}

\begin{document}

\title{Unsupervised feature learning from finite data by message passing: discontinuous versus continuous phase transition}

\author{Haiping Huang}
\affiliation{RIKEN Brain Science Institute, Wako-shi, Saitama
351-0198, Japan}
\author{Taro Toyoizumi}
 \affiliation{RIKEN Brain
Science Institute, Wako-shi, Saitama 351-0198, Japan}
\date{\today}

\begin{abstract}
Unsupervised neural network learning extracts hidden features from
unlabeled training data. This is used as a pretraining step for
further supervised learning in deep networks. Hence, understanding
unsupervised learning is of fundamental importance. Here, we study the unsupervised learning from a finite
number of data, based on the restricted Boltzmann machine learning. Our study inspires an efficient message passing
algorithm to infer the hidden feature, and estimate the entropy of
candidate features consistent with the data. Our analysis reveals that the learning requires only a few data if
the feature is salient and extensively many if the feature is weak. Moreover,
the entropy of candidate features monotonically decreases with data
size and becomes negative (i.e., entropy crisis) before the message passing becomes unstable, suggesting a discontinuous phase transition.
In terms of convergence time of the message passing algorithm, the unsupervised learning exhibits an
easy-hard-easy phenomenon as the training data size increases.  All
these properties are reproduced in an approximate Hopfield model,
with an exception that the entropy crisis is absent, and only continuous phase transition is observed. This key difference is also confirmed in a handwritten digits dataset. This study deepens our understanding of
unsupervised learning from a finite number of data, and may
provide insights into its role in training deep networks.

\end{abstract}

\pacs{02.50.Tt, 87.19.L-, 75.10.Nr}
 \maketitle

Humans and other animals can learn new concepts from only a handful of training data but standard machine learning algorithms require 
huge data to uncover hidden features~\cite{Lake-2015}. Learning hidden features in unlabeled data is called
unsupervised learning. How many data are required to learn a feature? What key factors determine the success of unsupervised learning?
These fundamental questions are largely unsolved so far, and rarely explained by a physics model. Understanding how data size confines learning process 
is a topic of interest not only in machine
learning~\cite{Hinton-2007} but also in cognitive
neuroscience~\cite{Kersten-2004}. The underlying neural mechanism or inspired
algorithms are still elusive, but recent progress in mean field theory of restricted Boltzmann machine~\cite{Huang-2015b} allows us to develop
a statistical mechanics model to understand how learning improves with data size.

Restricted Boltzmann machine (RBM) is a basic block widely used in building a deep belief network~\cite{Hinton-2006a,Bengio-2013}. It consists of two layers of neurons.
The visible layer receives data,
and based on this, the hidden layer builds an internal representation. No lateral connections exist within each
layer for computational efficiency. The symmetric connections between visible and hidden neurons represent
hidden features in the data that the network learns. A common
strategy to train a RBM is to apply a gradient-descent update
algorithm of the connections based on stochastically sampled neural activities~\cite{Bengio-2013}. However, no theory was proposed to
address how learning improves and how the number of candidate features decreases with data size.

Here we tackle this problem by taking a different perspective. We propose a Bayesian inference framework to uncover the connections, i.e., a feature vector, from data by explicitly modeling its posterior
distribution. Based on this framework, we develop a message passing algorithm for inferring
an optimal feature vector, while monitoring its entropy 
to quantify how many candidate feature vectors are consistent with
given data.

In this study, we consider a RBM~\cite{Hinton-2006b,Huang-2015b} with a single hidden neuron, whose
activity is generated according to $P(\bs,h)\propto e^{-\beta
  E(\bs,h)/\sqrt{N}}$, where $\bs=\{\sigma_i| \sigma_i=\pm 1,
i=1,\dots,N\}$ represents binary activity of $N$ visible neurons,
$h=\pm 1$ is the activity of the hidden neuron, $\beta/\sqrt{N}$ is the system-size and inverse-temperature dependent scaling factor, and $E(\bs,h)=-h \bx^{\rm T} \bs$ is the energy function
characterized by a feature vector $\bx$ (${\rm ^T}$ indicates vector
transpose). We assume that each element of the feature
vector takes a binary value $\xi_i=\pm 1$.  While generalization to a case with multiple hidden
neurons is possible, we do not explore it here as the analysis
becomes more involved. 

To perform unsupervised learning, we generate $M$ independent samples as training data $\{\bs^a\}_{a=1}^M$ from a
RBM with randomly generated true feature vector $\bx^{\rm true}$, where each element
is drawn from $\pm 1$ with equal probability.
Another RBM learns this feature
vector from the data.
 We formulate the learning process as Bayesian
inference.  Given the training data, the posterior
distribution of the feature vector is
\begin{equation}\label{Pobs}
 P(\bx|\{\bs^{a}\}_{a=1}^{M})\propto\prod_{a=1}^{M}P(\bs^{a}|\bx)=\frac{1}{Z}\prod_{a=1}^{M}\cosh\left(\frac{\beta}{\sqrt{N}}\bx^{{\rm T}}\bs^{a}\right),
\end{equation}
where $Z$ is the partition function of the model and we assume
a uniform prior about $\bx$. Here, the data $\{\bs^{a}\}_{a=1}^{M}$ serves as the quenched disorder (data constraints), and the inverse-temperature parameter $\beta$ 
characterizes the learning difficulty on a network of dimension $N$. If $M>1$, the model becomes non-trivial as the partition function can not be computed exactly for a large $N$. This $M$ can be proportional to the system size, and in this case we define
a data density as $\alpha=M/N$. Hereafter, we omit the conditional dependence of $P(\bx|\{\bs^{a}\}_{a=1}^{M})$ on $\{\bs^{a}\}_{a=1}^{M}$.

\begin{figure}
\centering
    \includegraphics[bb=46 527 418 736,scale=0.55]{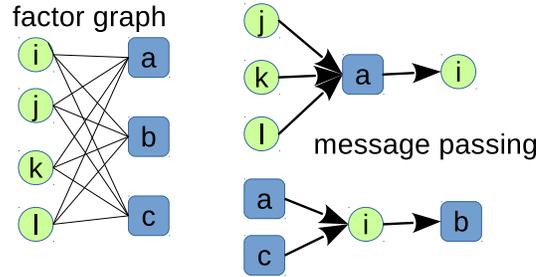}
  \caption{ (Color online) Schematic illustration of factor graph representation and message passing. Left panel: circle nodes indicate features to be
  inferred. Square nodes indicate data constraints. A connection $\sigma_i^{a}$ indicates how the feature $\xi_i$ is related to $a$-th data. 
  Right panel: the top panel shows data node $a$ collects messages from its neighboring features other than node $i$ and produces an output message to node $i$. The bottom panel shows
  feature
  node $i$ collects messages from its neighboring data nodes other than $b$ and produces an output message to data node $b$. Iteration of these messages gives a coherent
  understanding of the feature learning process.
  }\label{rbm}
\end{figure}

In the following, we compute the maximizer of the posterior marginals (MPM) estimator $\hat\xi_i = \arg\max_{\xi_i}P_i(\xi_i)$~\cite{Nishimori-2001}, which maximizes the overlap
$q=\frac{1}{N}\sum_{i=1}^N \xi^{\rm true}_i \hat\xi_i$ between the
true and estimated feature vectors. If $q=0$, the data do not give any information about the feature vector. If $q=1$, the feature vector is perfectly estimated.
Hence, the task now is to compute marginal probabilities, e.g., $P_i(\xi_i)$, which is still a hard problem due to the interaction among data constraints. However, by mapping
the model (Eq.~(\ref{Pobs})) onto a factor graph~\cite{MM-2009,Huang-2015b}, the marginal probability can be estimated by message passing (Fig.~\ref{rbm}) as we explain below.
We first assume that elements of the feature vector on the factor graph are weakly correlated (also named Bethe approximation~\cite{cavity-2001}), then by using the cavity method~\cite{MM-2009}, we define a cavity
probability $P_{i\rightarrow a}(\xi_i)$ of $\xi_i$ on a modified factor graph with data node $a$ removed. Under the weak correlation assumption, $P_{i\rightarrow a}(\xi_i)$
satisfies a recursive equation (namely belief propagation (BP) in computer science~\cite{Yedidia-2005}):
\begin{subequations}\label{bp0}
\begin{align}
P_{i\rightarrow a}(\xi_{i})&\propto
\prod_{b\in\partial i\backslash
a}\mu_{b\rightarrow i}(\xi_{i}),\label{bp2}\\
\begin{split}
\mu_{b\rightarrow i}(\xi_{i})&=\sum_{\{\xi_{j}|j\in\partial
  b\backslash
  i\}}\cosh\left(\frac{\beta}{\sqrt{N}}\bx^{{\rm T}}\boldsymbol{\sigma}^{b}\right)\prod_{j\in\partial
  b\backslash i}P_{j\rightarrow b}(\xi_{j}),\label{bp1}
\end{split}
\end{align}
\end{subequations}
where the proportionality symbol $\propto$ omits a normalization constant,
$\partial i\backslash a$ defines the neighbors of feature node $i$ except
data node $a$, $\partial b\backslash i$ defines the neighbors of
data node $b$ except feature node $i$, and the auxiliary quantity
$\mu_{b\rightarrow i}(\xi_i)$ represents the contribution from
data node $b$ to feature node $i$ given the value of
$\xi_i$~\cite{MM-2009}. An equation similar to Eq.~(\ref{bp0}) was recently derived
to compute activity statistics of a RBM~\cite{Huang-2015b}.

In the thermodynamic limit, the sum inside the hyperbolic cosine function excluding the $i$-dependent term in Eq.~(\ref{bp1}) is a random variable following a normal distribution with mean $G_{b\rightarrow i}$ and variance
$\Xi_{b\rightarrow i}^{2}$~\cite{Huang-2015b}, where $G_{b\rightarrow
i}=\frac{1}{\sqrt{N}}\sum_{j\in\partial b\backslash i}\sigma_{j}^{b}m_{j\rightarrow b}$ and
$\Xi^{2}_{b\rightarrow i}\simeq\frac{1}{N}\sum_{j\in\partial b\backslash
i}(1-m_{j\rightarrow b}^{2})$. The cavity magnetization is defined as $m_{j\rightarrow b}=\sum_{\xi_j}\xi_jP_{j\rightarrow b}(\xi_j)$. Thus the intractable sum over
all $\xi_j$ ($j\neq i$) can be replaced by an integral over the normal distribution. Furthermore, because $\xi_i$ is a binary variable, $P_{i\rightarrow a}(\xi_i)$ and
$\mu_{b\rightarrow i}(\xi_i)$ are completely characterized~\cite{Huang-2015b} by the cavity magnetization $m_{i\rightarrow a}$ and cavity bias $u_{b\rightarrow i}=\frac{1}{2}\ln\frac{\mu_{b\rightarrow i}(\xi_i=1)}{\mu_{b\rightarrow i}(\xi_i=-1)}$, respectively.
Using these expressions,
the BP equation (Eq.~(\ref{bp0})) could be reduced to the following practical recursive equations:
\begin{subequations}\label{bp3}
\begin{align}
m_{i\rightarrow a}&=\tanh\left(\sum_{b\in\partial i\backslash
a}u_{b\rightarrow i}\right),\\
u_{b\rightarrow i}&=\tanh^{-1}\left(\tanh(\beta G_{b\rightarrow i})\tanh(\beta\sigma_i^{b}/\sqrt{N})\right),\label{bp4}
\end{align}
\end{subequations}
where $m_{i\rightarrow a}$ can be interpreted as the message passing from feature $i$ to the data constraint $a$, while $u_{b\rightarrow i}$ can be
interpreted as the message passing from data constraint $b$ to feature $i$ (Fig.~\ref{rbm}). If the weak correlation assumption is
self-consistent, the BP would converge to a fixed point corresponding to a stationary point of the Bethe free energy function with respect to the cavity messages $\{m_{i\rightarrow a},u_{a\rightarrow i}\}$~\cite{MM-2009}.

By initializing the message on each link of the
factor graph (Fig.~\ref{rbm}), we run the BP equation
(Eq.~(\ref{bp3})) until it converges within a prefixed precision. From this fixed point, one can extract useful information about the true feature vector, by calculating the marginal probability as
$P_i(\xi_i)=\frac{1+m_i\xi_i}{2}$ where $m_i=\tanh\left(\sum_{b\in\partial i}u_{b\rightarrow i}\right)$. An alternative strategy to use passing messages to infer the true
feature vector is called reinforced BP (rBP). This strategy is a kind of soft-decimation~\cite{Zecchina-2006}, which progressively enhances/weakens the current 
local field ($h_i^{t}=\sum_{b\in\partial i}u_{b\rightarrow i}$) each feature component feels by the reinforcement rule $h_i^t\leftarrow h_{i}^{t}+h_{i}^{t-1}$ with a probability $1-\gamma^t$, until
a solution ($\{\hat{\xi}_i={\rm sgn}(m_i)\}$) is stable over iterations.
$\gamma$ usually takes a value close to $1$.

Another important quantity is the number of candidate feature vectors consistent with the data, characterized by the entropy per neuron
$s=-\frac{1}{N}\sum_{\bx}P(\bx)\ln P(\bx)$. Under the Bethe approximation, $s$ is evaluated as summing up contributions from single feature nodes and data nodes
~\cite{SM}.

\begin{figure}
\centering
          \includegraphics[bb=60 8 716 522,scale=0.3]{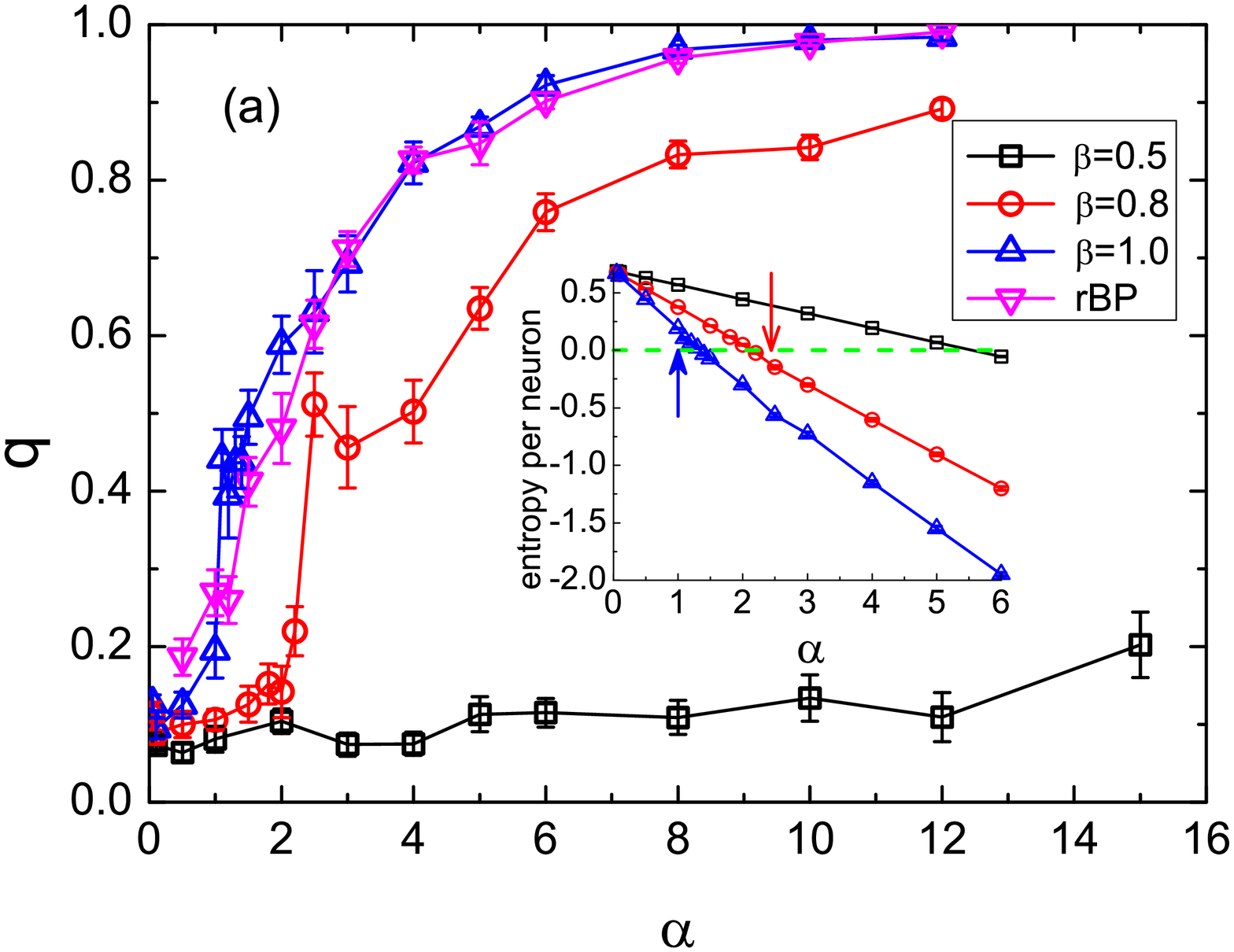}
     \hskip .05cm
     \includegraphics[bb=58 16 717 518,scale=0.3]{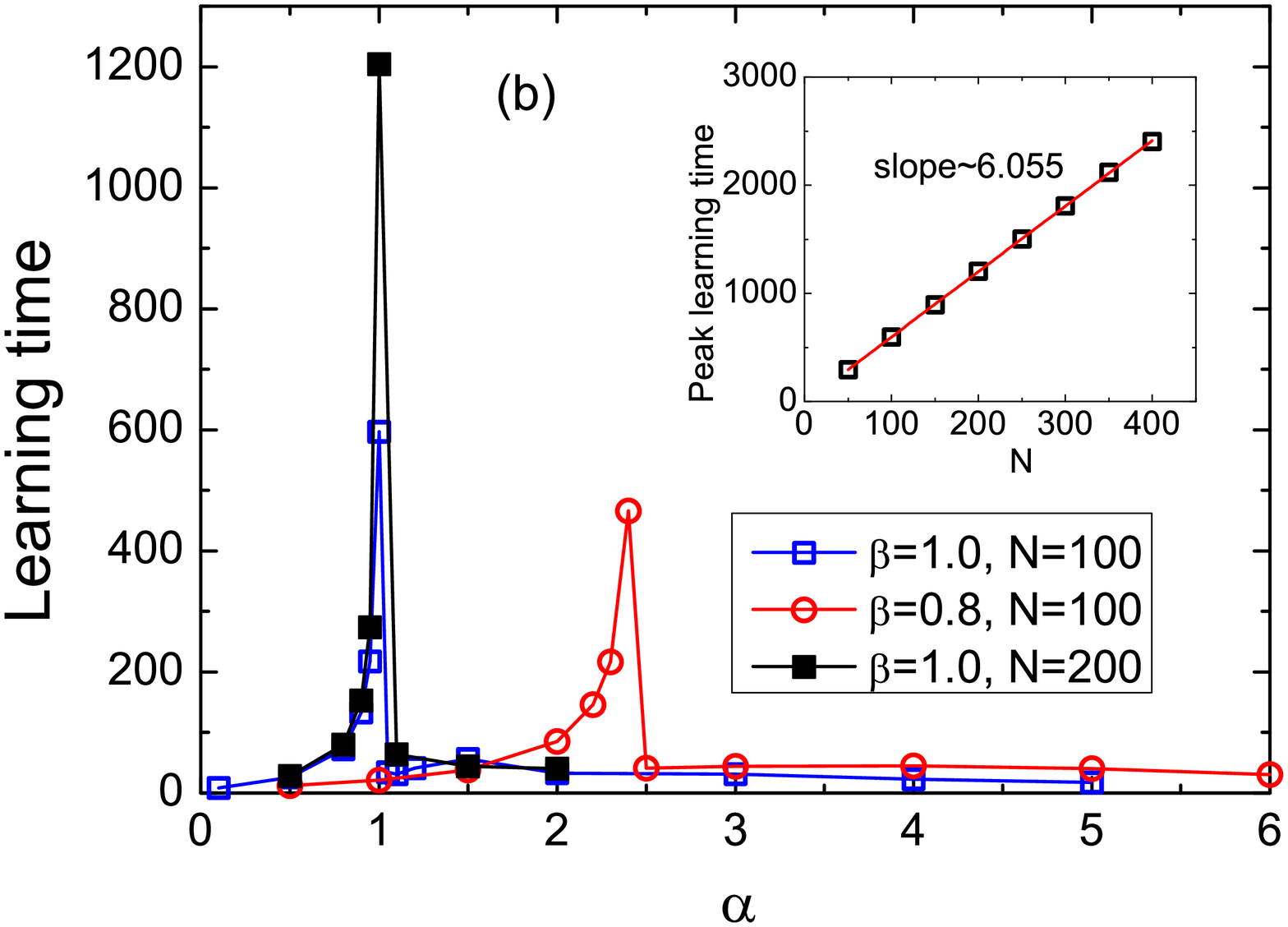}
  \caption{(Color online) Bayesian learning in the RBM model (Eq.~(\ref{Pobs})). $30$ random realizations of the model are considered. The error bars
  characterize the standard deviation. $N=100$. (a) Inference overlap $q$ versus data density $\alpha$ with
  different values of feature saliency $\beta$. Results obtained by rBP ($\gamma=0.95$) are also shown for comparison ($\beta=1$). The inset shows the entropy per neuron. Arrows in the inset indicate thresholds for continuous transitions.
  (b) Median of convergence (learning) time defined by the number of iterations at which BP converges to a fixed point.
  Two different values of $\beta$ are considered. For $\beta=1$, results for a larger $N$ are also shown. The inset shows that the peak learning time scales linearly with $N$ ($\beta=1$).
      }\label{ori}
 \end{figure}

We use the above mean field theory to study unsupervised learning of
the feature vector in single random realizations of the true feature vector and analyze how its
performance depends on model parameters. We first explore effects of $\beta$ ($N=100$). Note that $\beta$ scales the energy and hence tunes the
difficulty of the learning. As shown in
Fig.~\ref{ori} (a), it requires an extensive number of data to
learn a weak feature vector ($\beta=0.5$). However, as $\beta$ increases, the inference
becomes much better. The overlap grows more rapidly
at $\beta=1.0$. When $\alpha$ is above
$10$, one can get a nearly perfect inference of the feature vector. We also use the reinforcement strategy to infer the true feature vector, and it has nearly the same
performance with reduced computer time, because the estimation of the true feature need not be carried out at the fixed point.

Remarkably, the overlap improves at some $\alpha$, revealing that the RBM could extract the
hidden feature vector only after sufficient data are shown. This critical $\alpha$ decreases with
the saliency $\beta$ of the hidden feature. A statistical analysis of Eq.~(\ref{bp3}) reveals that $\alpha_c=\frac{1}{\beta^4}$~\cite{SM}. Next, we show the entropy per neuron in the inset of Fig.~\ref{ori} (a).
This quantity that
describes how many feature vectors are consistent with the data
becomes negative (i.e., entropy crisis~\cite{REM-1980,Mezard-2004})
at a zero-entropy $\alpha_{{\rm ZE}}$. However, the BP equation is still stable (convergent),
and thus the instability occurs after the entropy crisis. This
suggests the existence of a discontinuous glass transition at a value of $\alpha$ less than or equal to $\alpha_{{\rm ZE}}$, as commonly observed in some spin
glass models of combinatorial satisfaction problems~\cite{Montanari-2001,Mezard-2004,Huang-2015a,Huang-2014}.
As shown in the inset of Fig.~\ref{ori} (a), $\alpha_c$ can be either larger or smaller than $\alpha_{{\rm ZE}}$, depending on $\beta$. If $\alpha_c<\alpha_{{\rm ZE}}$, 
a continuous transition is followed by a discontinuous transition, where intra-state and inter-state overlaps~\cite{Barkai-1990} bifurcate discontinuously.
If $\alpha_c>\alpha_{{\rm ZE}}$, the predicted continuous transition is inaccurate at least under the replica symmetric assumption. Detailed discussions about this
novel property will be provided in a forthcoming extended work~\cite{Huang-2016b}.
Interestingly, despite the likely glass transition and
entropy crisis, the fixed point of BP still yields good inference of
the feature vector, which may be related to the Nishimori condition (Bayes-optimal inference)~\cite{Iba-1999,Huang-2016b}, since the temperature parameter used in inference is the same as
that used to generate the data.

Next, we study the median of learning time. The learning time is measured as the number of iterations when the message passing converges. 
Fig.~\ref{ori} (b) shows that learning is fast at
small $\alpha$, slow around the critical $\alpha$, and becomes
fast again at large $\alpha$. This easy-hard-easy phenomenon can be understood as follows. When a few data are presented, the inference is less constrained, and thus there exist many candidate feature vectors consistent with the data, 
the BP converges fast to estimate the marginals.
Once relatively many data are presented, the number of candidate feature vectors reduces (Fig.~\ref{ori} (a)), and the BP requires more iterations to
find a candidate feature vector. We also observe that the peak learning time at $\alpha_c$ scales linearly with $N$ within the
measured range (the inset of Fig.~\ref{ori} (b)). This intermediate region is thus termed hard phase.
Large $\beta$ moves the hard phase to the small $\alpha$ region.
Once the number of data is sufficiently large, the inference becomes
easy once again as the number of necessary iterations drops drastically. This may be because the feature space around the true feature vector dominates the posterior
probability in the presence of sufficient data.

It is interesting to show that one can also perform the same unsupervised learning by using an associative memory (Hopfield) model defined by
\begin{equation}\label{Hopf}
 \tilde{P}(\bx)\propto\prod_{a}e^{\frac{\hb}{2N}\Bigl(\bx^{{\rm T}}\bs^{a}\Bigr)^2},
\end{equation}
where $\hb=\beta^2$. This posterior distribution about a feature vector $\bx$ given data $\{\bs^{a}\}_{a=1}^{M}$ can be obtained by a small-$\beta$ expansion of Eq.~(\ref{Pobs})~\cite{Barra-2012}.
This relationship indicates that
one can infer the feature vector of a RBM using this Hopfield
model if $\beta$ is small enough. In this case, the true feature vector is also interpreted as
a unique memory pattern in the Hopfield model. By a small-$\beta$ expansion, we interpret the unsupervised learning in a 
RBM model as recovering stored pattern in a Hopfield model from noisy data~\cite{Monasson-2011}. 
In the following, we generate data by a RBM with true feature $\bx^{\rm true}$ and compute the MPM estimator $\hat{\bx}$ of
the memory pattern in Eq. (\ref{Hopf}) based on the given data.

Analogous to the derivation of Eq.~(\ref{bp3}), we can derive the practical BP corresponding to the posterior probability (Eq.~(\ref{Hopf})):
\begin{equation}\label{bp5}
 m_{i\rightarrow a}=\tanh\left(\frac{\hb}{\sqrt{N}}\sum_{b\in\partial i\backslash a}\sigma_i^{b}\hG_{b\rightarrow i}F_{b\rightarrow i}\right),
\end{equation}
where $\hG_{b\rightarrow i}=\frac{1}{\sqrt{N}}\sum_{k\in\partial b\backslash i}\sigma_k^{b}m_{k\rightarrow b}$, $F_{b\rightarrow i}=1+\frac{\hb C_{b\rightarrow i}}{1-\hb C_{b\rightarrow i}}$ in which
$C_{b\rightarrow i}=\frac{1}{N}\sum_{k\in\partial b\backslash i}(1-m_{k\rightarrow b}^2)$. The entropy of candidate feature vectors can also be evaluated from the fixed
point of this iterative equation~\cite{SM}.

\begin{figure}
\centering
    \includegraphics[bb=60 33 717 515,scale=0.35]{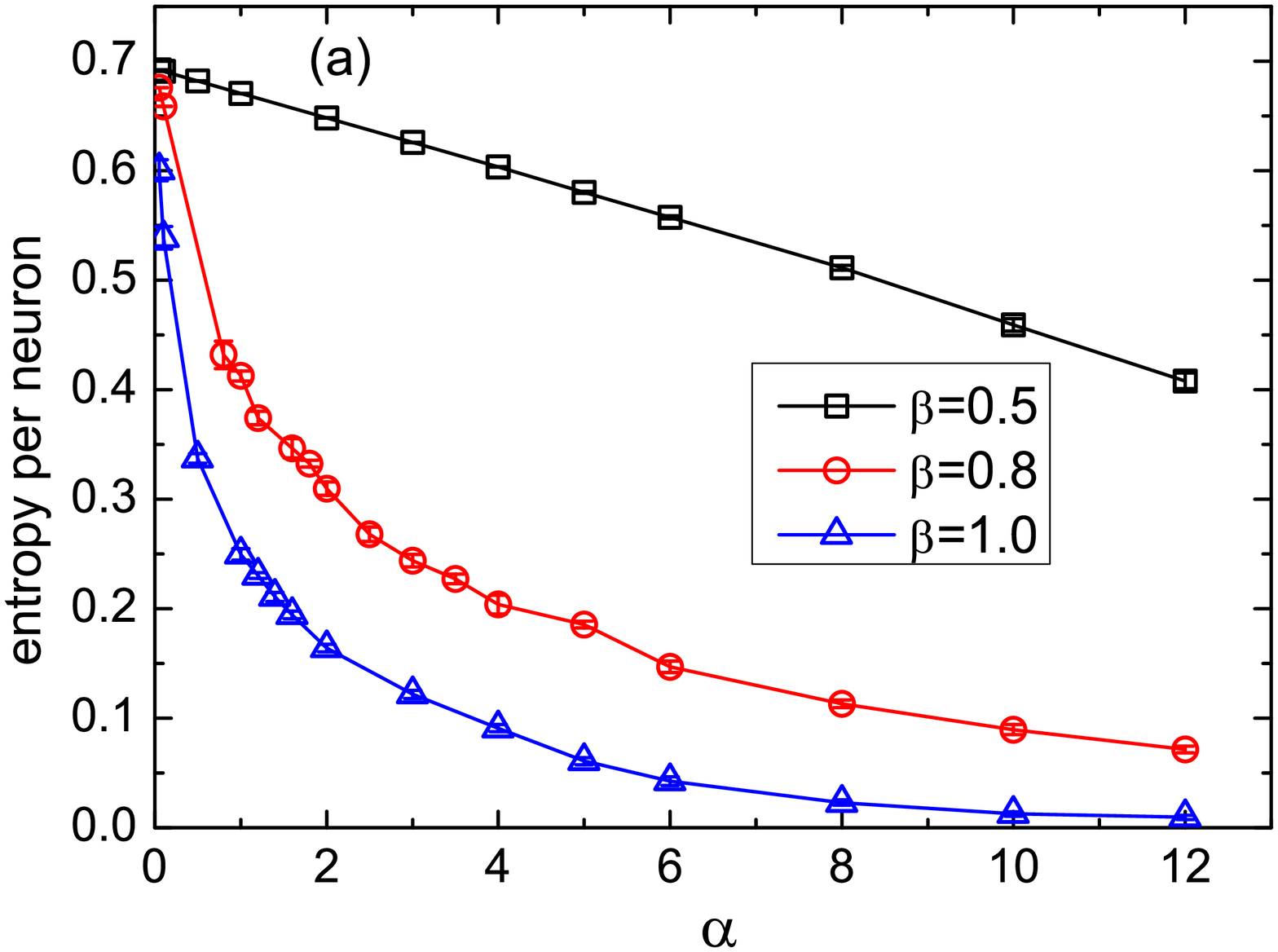}
   \vskip .05cm
     \includegraphics[bb=66 13 711 513,scale=0.35]{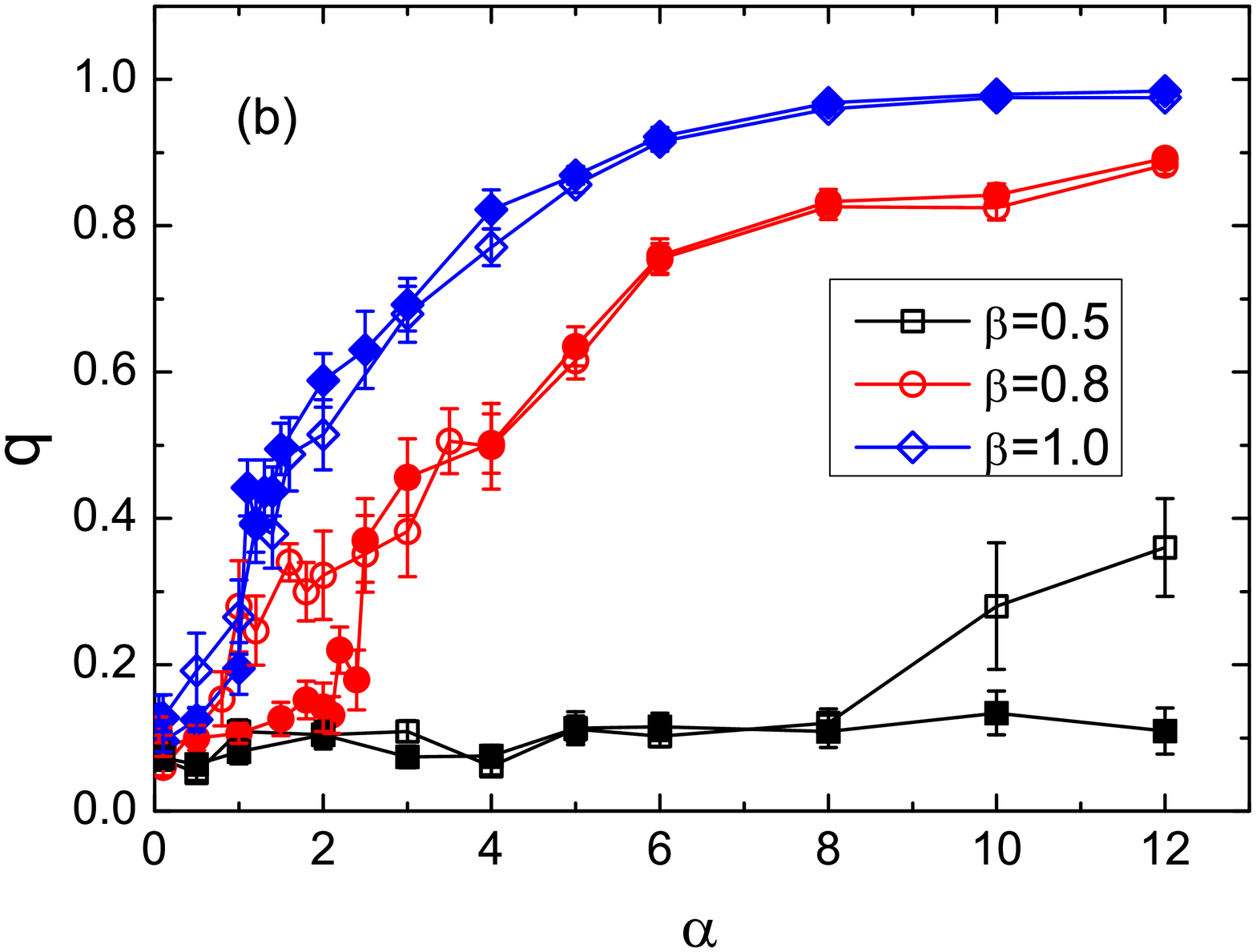}
  \caption{
  (Color online) Bayesian learning in the approximate Hopfield model (Eq.~(\ref{Hopf})). $30$ random realizations of the model are considered. The error bars
  characterize the standard deviation. $N=100$. (a) Entropy per neuron versus the data density $\alpha$. (b) Overlap versus the data density. Results for RBM (solid symbols)
  are shown for comparison.
  }\label{hopf}
\end{figure}

Bayesian learning performance of the Hopfield model is shown in Fig.~\ref{hopf}.
This model does not show an entropy crisis in the explored range of $\alpha$.
As $\alpha$ increases, the entropy decreases much more slowly for weak features than for strong ones. 
For $\beta=0.5$, the overlap stays slightly above
zero for a wide range of $\alpha$. At a sufficiently large $\alpha$
($\sim8$), the overlap starts to increase continuously. It is impossible to
predict underlying structures if the number of data is insufficient. This phenomenon is named
retarded learning first observed in unsupervised learning based on
Gaussian or mixture-of-Gaussian data~\cite{Biehl-1994,Nadal-1994}. At large $\alpha$ where the entropy value approaches zero, the overlap approaches one. 
All the properties except the entropy crisis are qualitatively similar in both the approximate Hopfield and RBM model. The absence of entropy crisis
may be related to the absence of $p$-spin ($p>2$) interanctions in the approximate model, which has thus only
the continuous glass transition at $\alpha_c=\left(\frac{1}{\hb}-1\right)^2$, predicted by a statistical analysis of Eq.~(\ref{bp5})~\cite{SM}. 
Note that the spin glass transition in a standard Hopfield model where many random patterns are stored is of second order~\cite{somp-1985}.
The current analysis sheds light on understanding the relationship between RBM and associative memory networks~\cite{Barra-2012,Mezard-2016}
within an unsupervised learning framework.

\begin{figure}
     \includegraphics[bb=60 9 713 531,scale=0.35]{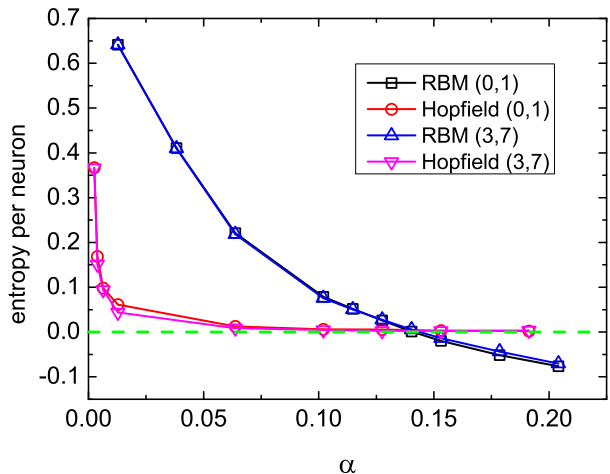}
  \caption{
  (Color online) Entropy per neuron versus $\alpha$ for MNIST
  dataset (handwritten digits ($0$,$1$) and ($3,7$)) at $\beta=1$.
  }\label{mnist}
\end{figure}

Finally, we test our theory on the MNIST handwritten digit
dataset~\cite{Lecun-1998}. For simplicity, we consider some combinations of two different digits
(e.g., $0$ and $1$, $3$ and $7$). Each digit is represented by a $28\times28$ gray-scale
image. Results show that the real dataset shares common properties
with our model (Fig.~\ref{mnist}), which does not change qualitatively when different combinations even the whole dataset are used. The inferred feature vector
(receptive field of hidden neuron) improves as the number of
data grows, serving as a local structure detector~\cite{SM}. This is indicated by the precision-recall curve moving to the rightmost upper corner of the plot as
the data size increases~\cite{SM}. Importantly, the presence and absence of the
entropy crisis in a RBM and a Hopfield model, respectively, are
also confirmed in this real dataset.

In conclusion, we build a physics model of unsupervised learning based on the RBM and propose a Bayesian inference framework to extract a hidden feature vector from
a finite number of data. The mean field theory inspires an
efficient message passing procedure to infer the hidden feature. Unlike previous approaches, each data in this work
is treated as a constraint on the factor graph, and the message passing carries out a probabilistic inference of the hidden feature without sampling activities of neurons.
We show that, salient features can be recovered by even a few data. Conversely, it is impossible to recover the weak features by a finite amount of data.
Interestingly, the entropy of candidate feature vectors becomes negative before the message passing algorithm
becomes unstable, suggesting a discontinuous glass transition to resolve the entropy crisis, a typical statistical mechanics phenomenon 
revealed in studies of spin glass models~\cite{Montanari-2001,Mezard-2004,Huang-2015a,Huang-2014}. 
In terms of the convergence time of the message passing algorithm, we reveal the easy-hard-easy phenomenon for the current unsupervised learning, 
which is explained by our theory. All these properties except the entropy crisis are also observed in an approximate Hopfield model,
where we infer from data a hidden feature of a RBM.
Remarkably, these phenomena
are also confirmed in a real dataset (Fig.~\ref{mnist}). This work provides a theoretical basis to understand efficient
neuromorphic implementation of RBM with simple binary feature elements~\cite{Bengio-2015}. We also hope our study will
provide important insights into a physics understanding of unsupervised learning, especially its
important role in pretraining deep neural networks for superior performances~\cite{Bengio-2013}.


\begin{acknowledgments}
 H.H. thanks Pan Zhang for useful discussions. This work was supported by the program for Brain Mapping by Integrated Neurotechnologies
for Disease Studies (Brain/MINDS) from Japan Agency for Medical Research and development, AMED, and by RIKEN Brain
Science Institute.
\end{acknowledgments}

\appendix
\newpage
\onecolumngrid
\setcounter{figure}{0}
\renewcommand{\thefigure}{S\arabic{figure}}
\renewcommand\theequation{S\arabic{equation}}
\setcounter{equation}{0}
\section{The number of candidate feature vectors in the RBM model}
For the unsupervised learning in the RBM model,  one important quantity is the number of candidate feature vectors consistent with the input noisy data, characterized by the entropy per neuron
$s=-\frac{1}{N}\sum_{\bx}P(\bx)\ln P(\bx)$. Under the Bethe approximation~\cite{cavity-2001}, $s$ is evaluated as summing up contributions from single feature nodes and data nodes:
\begin{equation}
 Ns=\sum_i\Delta S_i-(N-1)\sum_a\Delta S_{a},
\end{equation}
where single feature node contribution is expressed as $\Delta S_i=\sum_{a\in\partial i}\left[\beta^2\Xi_{a\rightarrow i}^2/2+
\ln\cosh(\beta G_{a\rightarrow i}+\beta\sigma_i^{a}/\sqrt{N})\right]+\ln\left(1+\prod_{a\in\partial i}\mathcal{G}_{a\rightarrow i}\right)-\left[\sum_{a\in\partial i}
\mathcal{H}_{a\rightarrow i}(+1)+\prod_{a\in\partial i}\mathcal{G}_{a\rightarrow i}\sum_{a\in\partial i}\mathcal{H}_{a\rightarrow i}(-1)\right]/\left(1+\prod_{a\in\partial i}\mathcal{G}_{a\rightarrow i}\right)$,
and single data node contribution $\Delta S_a=\ln\cosh(\beta G_a)-\beta^2\Xi_{a}^2/2-\beta G_a\tanh(\beta G_a)$. We define $\mathcal{G}_{a\rightarrow i}=e^{-2u_{a\rightarrow i}}$,
$\mathcal{H}_{a\rightarrow i}(\xi_i)=\beta^2\Xi_{a\rightarrow i}^2+(\beta G_{a\rightarrow i}+\beta\sigma_i^{a}\xi_i/\sqrt{N})\tanh(\beta G_{a\rightarrow i}+\beta\sigma_i^{a}\xi_i/\sqrt{N})$,
$G_a=\frac{1}{\sqrt{N}}\sum_{i\in\partial a}\sigma_i^{a}m_{i\rightarrow a}$, and $\Xi^2_a=\frac{1}{N}\sum_{i\in\partial a}(1-m^2_{i\rightarrow a})$.

\section{The number of candidate feature vectors in the approximate Hopfield model}
For the approximate Hopfield model, the entropy can be evaluated as
$Ns=\sum_i\Delta S_i-(N-1)\sum_a\Delta S_{a}$, where single feature node contribution reads $\Delta S_i=-\sum_{a\in\partial i}\left[\frac{1}{2}\ln(1-\hb C_{a\rightarrow i})+\frac{\hb C_{a\rightarrow i}}{2(1-\hb C_{a\rightarrow i})}
+\frac{\hb}{2}(1/N+\hG^{2}_{a\rightarrow i})F'_{a\rightarrow i}\right]+\ln\left(2\cosh(\hb H_i)\right)-(\hb H_i+\hb H'_i)\tanh(\hb H_i)$,
and single data node contribution $\Delta S_a=-\frac{1}{2}\ln(1-\hb C_{a})-\frac{\hb C_a}{2(1-\hb C_a)}-\frac{\hb}{2}\hG_a^2F'_a$, where $\hG_{a}=\frac{1}{\sqrt{N}}\sum_{k\in\partial a}\sigma_k^{a}m_{k\rightarrow a}$,
$F'_{a\rightarrow i}=\frac{\hb C_{a\rightarrow i}}{(1-\hb C_{a\rightarrow i})^2}$,
$F'_{a}=\frac{\hb C_{a}}{(1-\hb C_{a})^2}$, $C_a=\frac{1}{N}\sum_{k\in\partial a}(1-m_{k\rightarrow a}^2)$, $H_i=\frac{1}{\sqrt{N}}\sum_{b\in\partial i}\sigma_i^{b}\hG_{b\rightarrow i}F_{b\rightarrow i}$,
and $H'_i=\frac{1}{\sqrt{N}}\sum_{b\in\partial i}\sigma_i^{b}\hG_{b\rightarrow i}F'_{b\rightarrow i}$.

\section{A statistical analysis of practical BP equations}
We first statistically analyze the practical BP equations (Eq. (3) in the main text) for the RBM. In a large $N$ limit, the cavity bias can be approximated as $u_{b\rightarrow i}\simeq\frac{\beta\sigma_i^b}{\sqrt{N}}\tanh\beta G_{b\rightarrow i}$.
We then define a cavity field as $h_{i\rightarrow a}=\sum_{b\in\partial i\backslash a}u_{b\rightarrow i}$. The sum in the cavity field involves an order of $\mathcal{O}(N)$ terms, which are assumed to be nearly independent under
the replica symmetric assumption. Therefore, the cavity field follows a normal distribution with mean zero and variance $\alpha\beta^2\hat{Q}$, where $\hat{Q}\equiv\left<\tanh^2\beta G_{b\rightarrow i}\right>$. Note
that $G_{b\rightarrow i}$ is also a random variable subject to a normal distribution with zero mean and variance $Q$. $Q$ is defined by $Q=\frac{1}{N}\sum_{i}m_i^2$. To derive the variance of $G_{b\rightarrow i}$,
we used $\frac{1}{N}\sum_{k\in\partial b\backslash i}m_{k\rightarrow b}^2\simeq Q$, which is reasonable in the large $N$ limit (thermodynamics limit). Finally, we arrive at the following thermodynamic
recursive equation:
\begin{subequations}\label{rbmRM}
\begin{align}
Q&=\int Dz\tanh^2\beta\sqrt{\alpha\hat{Q}}z,\\
\hat{Q}&=\int Dz\tanh^{2}\beta\sqrt{Q}z,
\end{align}
\end{subequations}
where $Dz=\frac{dze^{-z^2/2}}{\sqrt{2\pi}}$. Note that $Q=0$ is a stable solution of Eq.~(\ref{rbmRM}) only when $\alpha\leq\alpha_c=\frac{1}{\beta^4}$. The threshold can be obtained by a Taylor
expansion of Eq.~(\ref{rbmRM}) around $Q=0$.

Next, we perform a statistical analysis of the practical BP equation (Eq. (5) in the main text) for the approximate Hopfield model. Similarly, a cavity field defined by
$h_{i\rightarrow a}=\frac{1}{\sqrt{N}}\sum_{b\in\partial i\backslash a}\sigma_{i}^{b}\hG_{b\rightarrow i}F_{b\rightarrow i}$ can be approximated by a random variable following a 
normal distribution with mean zero and variance $\frac{\alpha Q}{(1-\hb(1-Q))^2}$, where $Q\equiv\frac{1}{N}\sum_{i}m_i^2$. Consequently, we obtain the final thermodynamic equation as:
\begin{equation}\label{hopfRM}
 Q=\int Dz\tanh^2\left(\frac{\hb}{1-\hb(1-Q)}\sqrt{\alpha Q}z\right).
\end{equation}
Obviously, $Q=0$ is a solution of Eq.~(\ref{hopfRM}), which is stable up to $\alpha_c=\left(\frac{1}{\hb}-1\right)^2$. The threshold can be analogously derived by a linear stability 
analysis.

\section{Applications of our theory on MNIST dataset}
\label{methods}

\begin{figure}
          \includegraphics[bb=88 30 749 521,scale=0.3]{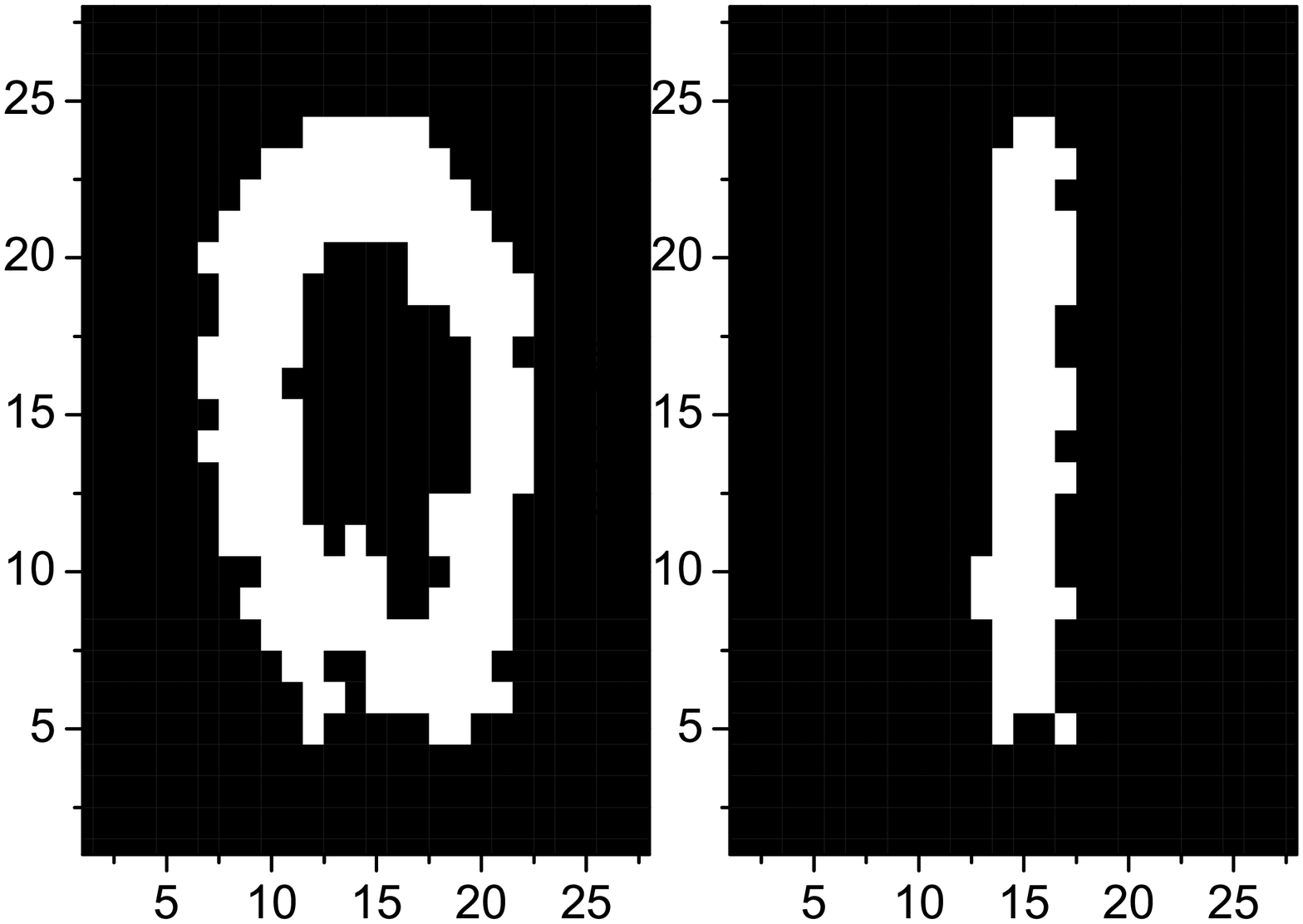}
  \caption{
 Two examples of digits ($0$ and $1$) from the MNIST dataset are
 shown.
     }\label{digit}
 \end{figure}

\begin{figure}
          \includegraphics[bb=92 37 740 550,scale=0.5]{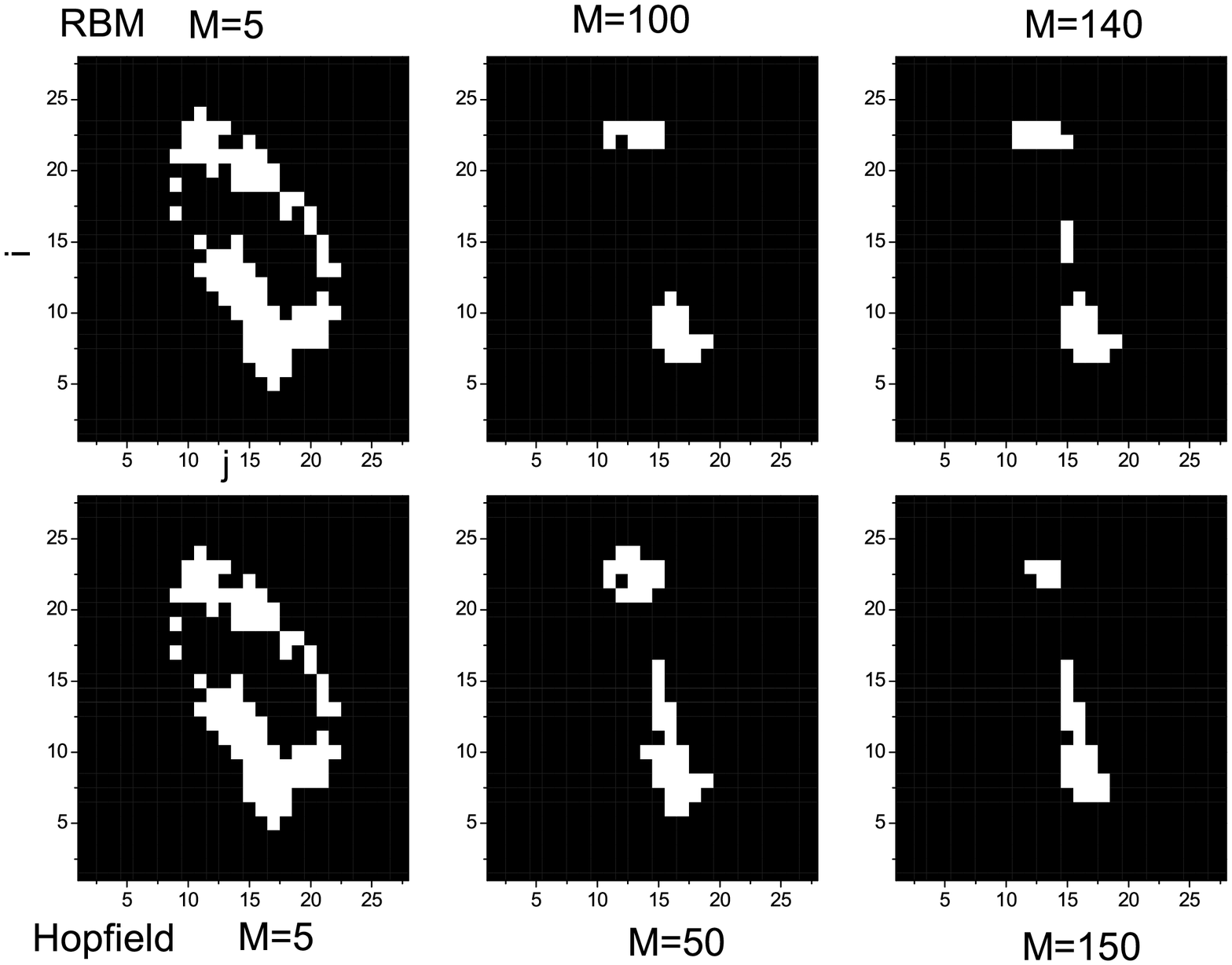}
  \caption{
 Inferred features evolve with the number of input images ($M$) on the MNIST
 dataset. The top panel displays the result for the RBM model, and the
 bottom panel displays the result for the approximate Hopfield model.
     }\label{RF}
 \end{figure}

We are interested in whether our model results hold in real dataset
or not. For simplicity, we consider MNIST dataset~\cite{Lecun-1998}
with only handwritten digits $0$ and $1$ (see Fig.~\ref{digit}).
Each digit is represented by a $28\times28$ gray-scale image. Given
the dataset (a set of digit $0$ and $1$ images), our theory could
estimate the corresponding hidden feature vector consistent with
these images. The feature vector is also called the receptive
field of the hidden neuron, because it determines the firing response of
the hidden neuron. We organize the inferred feature vector into a
$28\times28$ matrix (the same size as the input image), and plot the
matrix as the gray-scale image (black indicates $\xi_i=-1$, and
white indicates $\xi_i=1$). As shown in Fig.~\ref{RF}, the inferred
feature vector gets much better as the number of input images grows,
serving as a local structure detector. This is indicated by a few active synaptic weights in the center of the image, where
local characteristics of handwritten digits are approximately captured when suffcient data samples are shown to the machine. In a deep network composed of many layers of stacked RBM, the low-level features (e.g.,
edges or contours) detected by the hidden neuron could be passed to
deeper layers, where high-level information (e.g., object identity) could be extracted~\cite{DiCarlo-2012}.

\begin{figure}
 \centering
          \includegraphics[bb=88 30 749 521,scale=0.5]{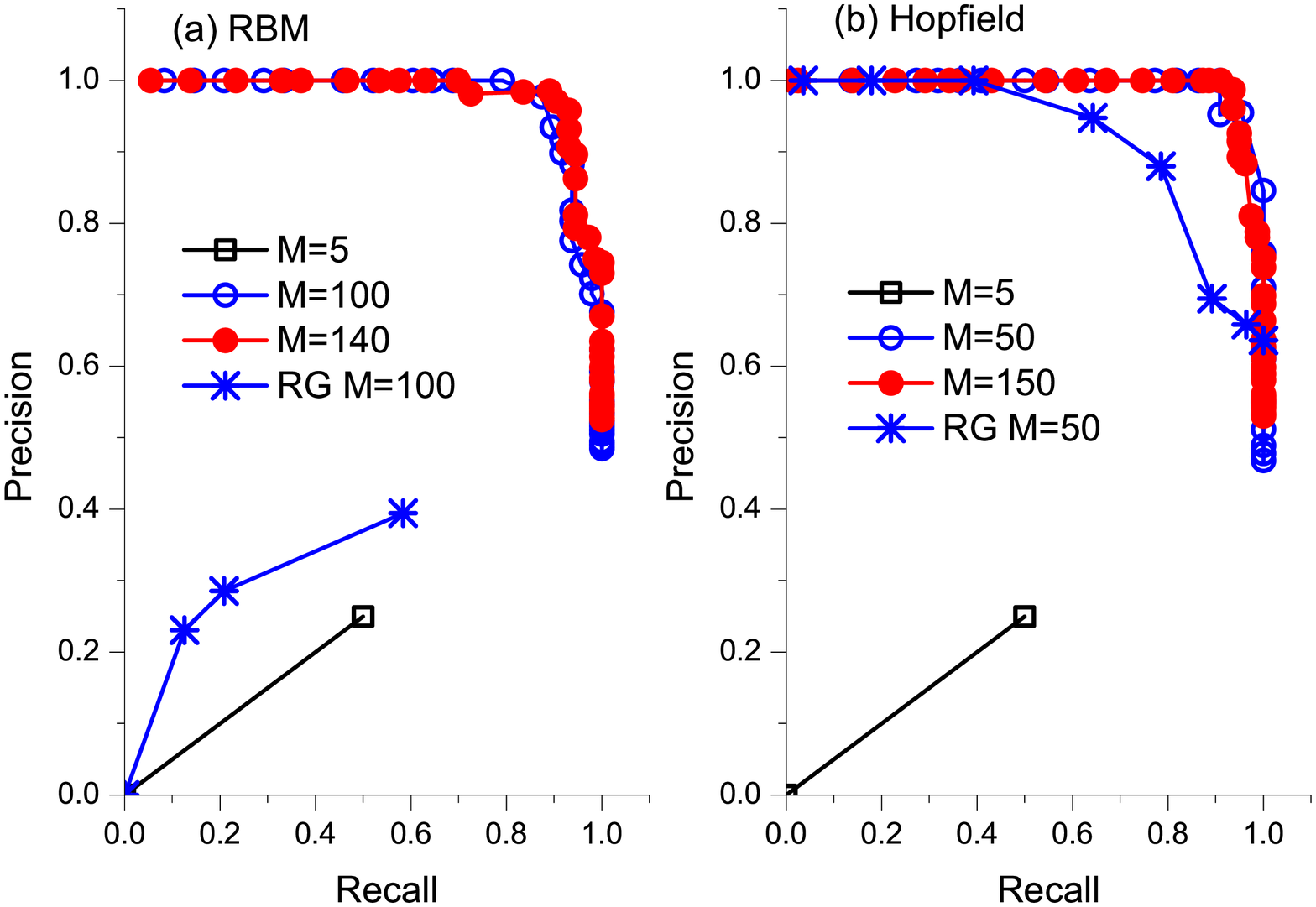}
  \caption{ (Color Online)
 The precision-recall curve corresponding to Fig.~\ref{RF}. The unsupervised learning performance is also compared with the random guess (RG) case.
     }\label{ROC}
 \end{figure}

 The effect of the data size on the feature learning performance can be quantitatively measured by the precision-recall curve. First, we computed the weighted sum (local field $H$) of the
hidden neuron given the input image. These fields are then ranked. Digits $1$ and $0$ are discriminated by introducing a threshold $H_{{\rm th}}$ for corresponding local fields.
The true positive (TP) event is identified when digit $1$ is predicted by a local field above the threshold. The false positive (FP) event is identified when digit $0$ is
wrongly predicted by a local field above the threshold. The false negative event (FN) is identified when digit $1$ is wrongly predicted by a local field below the threshold.
The recall (RC) is defined as ${\rm RC}=\frac{P_{{\rm TP}}}{P_{{\rm TP}}+P_{{\rm FN}}}$, and the precision (PR) is defined as ${\rm PR}=\frac{P_{{\rm TP}}}{P_{{\rm TP}}+P_{{\rm FP}}}$, where
$P_{{\rm TP}}$ is defined as the number of TP events in all presented samples. Thus the precision-recall curve is the parametric curve of ${\rm PR}(H_{{\rm th}})$ and ${\rm RC}(H_{{\rm th}})$.
The closer we can get to $(1,1)$ in the plot, the better the unsupervised feature learning understands the embedded feature structure of the data.
As shown in Fig.~\ref{ROC}, as the data size increases, the performance improves, and it behaves much better than a random guess of synaptic weights.

\end{document}